\documentclass{PoS}
\usepackage{subfig}

\title{Probing Backreaction Effects with Supernova Data}

\ShortTitle{Probing Backreaction Effects with Supernova Data}

\author{\speaker{Marina Seikel} and Dominik J. Schwarz
  \\
        Fakult\"at f\"ur Physik, Universit\"at Bielefeld, Postfach
        100131, 33501 Bielefeld, Germany\\
        E-mail: \email{mseikel@physik.uni-bielefeld.de}
        \email{dschwarz@physik.uni-bielefeld.de}}

\abstract{As the Einstein equations are non-linear,
spatial averaging and
temporal evolution do not commute. Therefore, the evolution of the
averaged universe is affected by inhomogeneities. It is, however,
highly controversial how large these cosmological backreaction effects
are. We use the supernova data of the Constitution set up to a
redshift of 0.1 in order to analyse to what extent the measurement of
the Hubble constant is affected. The size of the effect depends on the
size of the volume that is averaged over. The observational results
are then compared to the theory of the backreaction mechanism.
}

\FullConference{International Workshop on Cosmic Structure and Evolution\\
                 September 23-25, 2009\\
                 Bielefeld , Germany}

\begin{document}

\section{Introduction}
In cosmology it is very common to assume a homogeneous and isotropic
universe. As we know that there exist structures in the universe
\cite{structure}, this homogeneity and isotropy can only be
statistical on large scales \cite{homscale}, but not
exact. Due to the non-linearity of Einstein's equations spatial
averaging and temporal evolution do not commute. Thus, local
inhomogeneities can affect the expansion of the background universe
via the so-called backreaction mechanism
\cite{ellis,russ,buchert,backreaction}. 
In this work, we probe the influence of backreaction on the
measurement of the Hubble rate using supernova type Ia data.

\section{Averaging}
Many observables are averaged quantities. As all observed objects
lie on our past light cone, it would be appropriate to average the
observables over this light cone. This is however a very difficult
task that has not been achieved yet. Instead one can use spatial
averages at the cost of introducing 
an error. In order to keep this error at an acceptable level, it is
necessary to limit the use of spatial averages to low redshifts. In
our analysis we use supernovae up to $z=0.1$. In that range, this
averaging method is justified.

The averages of observables are calculated within a certain domain $D$.
Its volume is given by
\begin{equation}
V_D(t) \equiv \int_D W_D({\bf x}) \sqrt{\mbox{det}g_{ij}} \mbox{d}{\bf
  x} \;,
\end{equation}
where $W_D({\bf x})$ is the window function specifying the domain.
Then the spatial average of an observable $O$ within $D$ is
\begin{eqnarray}
\langle O \rangle_D\equiv \frac{1}{V_D(t)}\int_D W_D({\bf x}) O(t,{\bf x})
\sqrt{\mbox{det}g_{ij}}\mbox{d}{\bf x} \;.
\end{eqnarray}
An effective scale factor $a_D$ can be defined via the domain volume:
\begin{eqnarray}
\frac{a_D}{a_{D_0}}\equiv
\left(\frac{V_D}{V_{D_0}}\right)^{1/3} \;,
\end{eqnarray}
where the subscript 0 denotes the present time. The effective Hubble
rate then determined by
\begin{equation}
H_D\equiv\frac{\dot{a}_D}{a_D} \;.
\end{equation}

Following Buchert's formalism, the Einstein equations can be averaged
in order to obtain the effective Friedmann equations for a dust
universe \cite{buchert}:
\begin{eqnarray}
\left(\frac{\dot{a}_D}{a_D}\right)^2 &=& \frac{8\pi G}{3}\rho_{\rm
eff}\;,\\
 -\frac{\ddot{a}_D}{a_D} &=& \frac{4\pi G}{3}(\rho_{\rm
eff}+3p_{\rm eff}) \;.
\end{eqnarray}
These equation include the energy density and pressure of an effective
fluid, which are given by
 \begin{eqnarray}
\rho_{\rm eff} &\equiv& \langle \rho\rangle_D-\frac{1}{16\pi G}
\left(\langle Q\rangle_D+\langle {\cal R}\rangle_D\right)\;, \\
p_{\rm eff} &\equiv& - \frac{1}{16\pi G}\left(\langle Q\rangle_D-
\frac{1}{3}\langle {\cal R}\rangle_D\right)\;.
\end{eqnarray}
$\langle Q\rangle_D$ denotes the kinematical backreaction and $\langle
     {\cal R}\rangle_D$ the averaged spatial curvature.

We want to analyse the influence of backreaction effects on the
measurement of the Hubble rate. Its average value obtained by observing
objects within a domain $D$ is denoted as $H_D$. Assuming that there
exists a global value $H_0$, we can define the
fluctuation of the Hubble rate as
\begin{equation}
\delta_H \equiv \frac{H_D-H_0}{H_0} \;.
\end{equation}
Without backreaction the average value $\overline{\delta_H}$
equals zero. Considering backreaction effects, the value of 
$\overline{\delta_H}$ becomes slightly negative for small domain
sizes. But the main effect is that the variance of $\delta_H$ is
increased compared to the case when backreaction effects are not taken
into account.

\section{Gaussian window function}
\subsection{Method}
It is not obvious which choice of window function yields the best
results for a test of backreaction effects. 
The first try was to assume a spherically symmetric domain 
described by a gaussian window function
\begin{equation}\label{gaussian}
W_D(r)=\frac{1}{\sqrt{2\pi}R_D}\exp\left(-\frac{r^2}{2R_D^2}\right)\;,
\end{equation}
where $R_D$ specifies the size of the domain.
Then the variance of $\delta_H$ due to backreaction effects is given
by \cite{li}:
\begin{equation}\label{vargaussian}
{\rm Var}\left(\delta_H\right)=
\frac{25}{486\pi^3}\frac{1}{(1+z)^2}\left(\frac{R_{\rm H}}{R_D}\right)^4
 \int_0^{\infty}\!{\cal P}_{\varphi}(x/R_D)J^2_{3/2}(x) {\rm d}x \;,
\end{equation}
where $R_H$ is the Hubble radius. The values for the power spectrum
${\cal P}_{\varphi}$ are taken from WMAP5 measurements \cite{komatsu}.

For the analysis, we use supernova type Ia data from the Constitution
set \cite{hicken} up to a redshift of 0.1. We used the data that were
fitted with SALT2 \cite{guy}.
The number density of SNe
needs to be approximately constant within the considered domain. Thus,
the number of SNe $N(r)$ in the distance interval $[r,r+{\rm d}r]$
must be proportional to
$r^2W_D(r)$. That means that we have to choose a subset of SNe, whose
distribution in space corresponds to that of the considered window
function.

\begin{figure}
\centering
\subfloat[]{\includegraphics*[width=7.5cm]{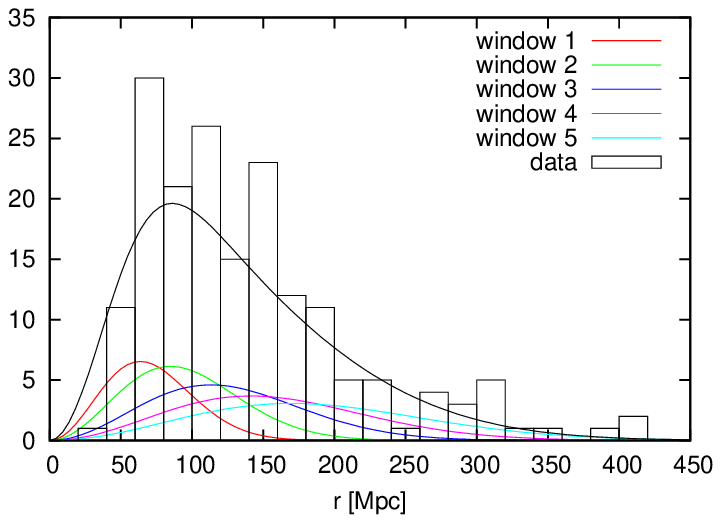}\label{gauss-bin}}
\hfill
\subfloat[]{\includegraphics*[width=7.5cm]{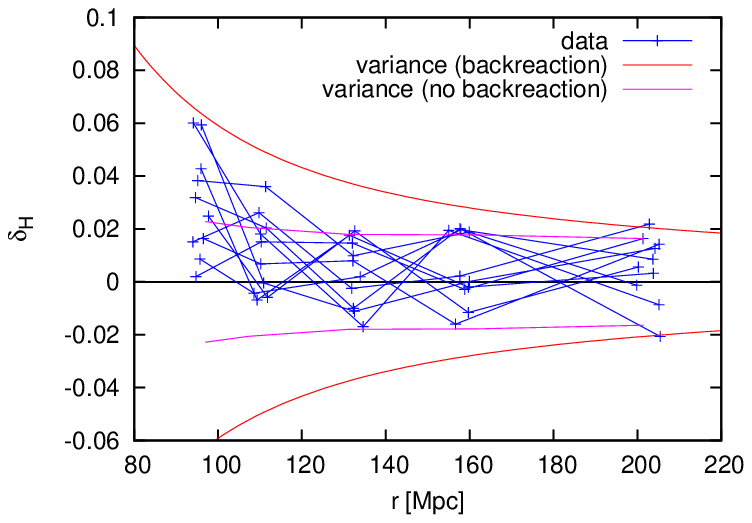}\label{gauss-result}}
\caption{Gaussian window function. (a) Distribution of the SNe 
  and $r^2W_D(r)$ for five different values of $R_D=45, 60, 80, 100,
  120$. The black curve is the sum of all five window functions
  $r^2W_D(r)$. (b) $\delta_H$ 
  obtained from ten different realisations of subsets. Also shown are
  the variances with and without backreaction.}
\end{figure}

Figure \ref{gauss-bin} shows the distribution of the 178 SNe in the
Constitution set up to redshift 0.1. For the test of backreaction
effects, it is essential to compare the value of $\delta_H$ of
differently sized domains. The domain size can be changed by varying
$R_D$ as given in equation (\ref{gaussian}). We chose to use five
domains. Their corresponding window functions $r^2W_D(r)$ are plotted
in the figure, the black curve being the sum of these functions.
However, the window functions only determine how many SNe at a certain
distance are assigned to a subset, but not which individual SNe. The
actual assignment of SNe to a subset within one realisation is done
randomly, but in a way that all subsets are disjoint and thus
statistically independent. As soon as one uses many realisations,
i.e. different assignments of SNe to the subsets, the statistical
independence is lost.

In order to calculate the Hubble rate $H_D$, we need to know the
redshift $z$ and the distance modulus $\mu$ for each supernova. The
distance modulus, however, depends on the calibration of the absolute
magnitude of the SNe. Thus, a different calibration leads to different
values of $\delta_H$, if one uses a global $H_0$ that was not obtained
by using the same data set, but by other observations such as WMAP. It
turned out that the test result is very sensitive to the
calibration. Therefore, we have to determine $H_0$ using the same data
set as for calculating $H_D$.

\subsection{Results}
Figure \ref{gauss-result} shows $\delta_H$ for ten random
realisations of subsets. The five data points obtained from the five
subsets in each realisation are connected by lines. $r$ is the average
distance of the SNe in a subset. The pink curves indicate purely the
measurement errors of the SNe. For the red curves $\sqrt{{\rm
    Var}\left(\delta_H\right)}$ from equation (\ref{vargaussian}) is
added in quadrature to the measurement errors. Here, the domain scale
$R_D$ has to be expressed in terms of the average distance as
$R_D=\sqrt{\pi/8}\, r$. The global Hubble rate that is needed for the
calculation of $\delta_H$ was chosen such that the data points at
large distances lie within the variance limits.
There is a trend of increasing $\delta_H$ with decreasing
distance. The model with backreaction effects seems to describe the
data better than the one without these effects.

In order to test backreaction effect, we need however a more
quantitative analysis. Therefore, we determined the optimal $H_0$ for
each realisation once with and once without backreaction effects. Then
we calculated the likelihoods of each model given the data. In 27 out
of 100 realisations the model including backreaction effects was
favoured. However, in none of the realisations one of the two models
was favoured significantly. 

\section{Tophat window function}
\begin{figure}
\centering
\subfloat[]{\includegraphics*[width=7.5cm]{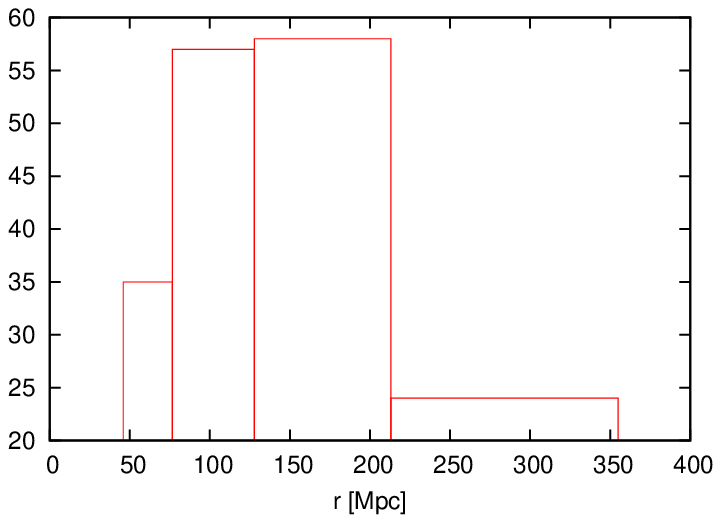}\label{tophat-bin}}
\hfill
\subfloat[]{\includegraphics*[width=7.5cm]{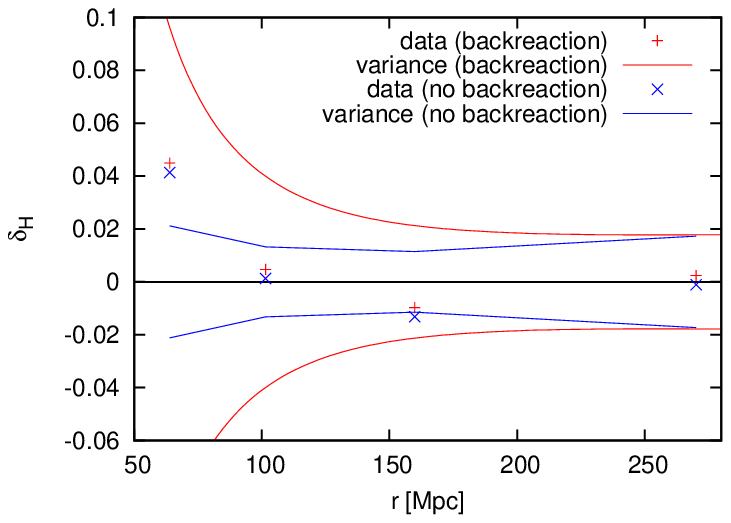}\label{tophat-result}}
\caption{Tophat window function. (a) Distribution of SNe.
  (b) $\delta_H$ obtained from the SN subsets. Also shown are
  the variances with and without backreaction.}
\end{figure}

As it was not possible to detect backreaction effects using a gaussian
window function, we tried a different ansatz. Since backreaction
effects are larger at smaller distances, our aim was to minimize the
distance of the first data point. This can be achieved by only using
the nearest SNe to calculate the first data point. So we binned the SN
data according to distance, where the binwidth is increased with
increasing distance (see figure \ref{tophat-bin}). The corresponding
window function is

\begin{equation}\label{tophat}
W_D(r)=\Theta(R_D) \Theta\left(r-\frac{5}{3}R_D\right) \;.
\end{equation}
Then the variance of $\delta_H$ is given by
\begin{equation}
{\rm Var}\left(\delta_H\right)=
\frac{2025}{153664\pi^2}\frac{1}{(1+z)^2}\left(\frac{R_{\rm H}}{R_D}\right)^4
 \int_0^{\infty}\!{\cal P}_{\varphi}(x/R_D)J^2_{3/2}(x) {\rm d}x\;.
\end{equation}
The relation between $R_D$ and the average distance is $R_D=49/68\,r$.

An advantage of the new window function (\ref{tophat}) is that the
assignment of SNe to subsets is unique. So we do not need to consider
different realisations. Like in the previous case, we determined the
optimal global Hubble rate $H_0$ for the model with and that without
backreaction and subsequently calculated $\delta_H$ for each
subset. The result is shown in figure \ref{tophat-result}. Note that
the data points for the two models differ slightly as we have used
different values of $H_0$. The quantitative analysis shows that the
model with backreaction is only about twice as likely as the models
without backreaction effects. Thus, we have not found any evidence for
backreaction.

\section{Conclusion}
Theoretically, backreaction influences the measurements
of the Hubble rate by increasing its variance. This effect should be
observed, if it was possible to measure the Hubble rate at different
locations in the universe. We are, however, restricted to our
local universe. So it is possible that our local measurements are by
chance consistent with a model that does not include backreaction
effects. If that was the case, we would not be able to detect those
effects using the test presented in this work. Therefore, the test can
potentially prove the existence of backreaction effects, but it cannot
prove that there are no such effects.

Using the currently available supernova data, we could find some
slight hint of backreaction, but no evidence. Larger data 
could help providing that evidence. A larger number of
supernovae leads to a smaller variance in the model without
backreaction. If the measured values of $\delta_H$ stayed
approximately the same for a sufficiently large data set, then the
data would lie significantly outside the variance limits
of a model without backreaction. In that way future data sets have the
potential of providing the evidence for the backreaction mechanism.

\acknowledgments
We thank David Hogg and William Press for discussions and
comments. The work of M.S. is supportet by the DFG under grant GRK
881.


\begin{thebibliography}{99}
\bibitem{structure} J.R.~Gott {\em et al.}, {\em A Map of the
  Universe}, {\em Astrophys.J.} {\bf 624} (2005) 463 [{\tt
    astro-ph/0310571}]
\bibitem{homscale} D.W.~Hogg {\em et al.}, {\em Cosmic homogeneity
  demonstrated with luminous red galaxies}, {\em Astrophys.J.} {\bf
  624} (2005) 54 [{\tt astro-ph/0411197}]; M.~Joyce {\em et al.}, {\em
  Basic properties of galaxy clustering in the light of recent results
  from the Sloan Digital Sky Survey}, {\em Astron.Astrophys.} {\bf
  443} (2005) 11 [{\tt astro-ph/0501583}]; D.J.~Schwarz, {\em Thoughts
  on the cosmological principle} in {\em Fundamental Interactions - A
  Memorial Volume for Wolfgang Kummer}, edited by D.~Grumiller,
  A.~Rebhan and D.~Vassilevich (World Scientific, Singapore, 2009)
  p.~267 [{tt arXiv:0905.0384}]
\bibitem{ellis} G.F.R.~Ellis, {\em Relativistic cosmology: Its nature,
  aims and problems} in {\em General relativity and gravitation},
  edited by B.~Bertotti, F.~de Felice and A.~Pascolini (Reidel,
  Dordrecht, 1984) p.~215
\bibitem{russ} H.~Russ, M.~Morita, M.~Kasai, G.~Boerner, {\em The
  Zel'dovich-type approximation for an inhomogeneous universe in
  general relativity: second-order solutions}, {\em Phys.Rev. D} {\bf
  53} (1996) 6881 [{\tt astro-ph/9512071}]; H.~Russ, M.H.~Soffel,
  M.~Kasai, G.~Boerner, {\em Age of the Universe: Influence of the
    Inhomogeneities on the global Expansion-Factor}, {\em Phys.Rev. D}
  {\bf 56} (1997) 2044 [{\tt astro-ph/9612218}]
\bibitem{buchert} T.~Buchert, {\em On average properties of
  inhomogeneous fluids in general relativity.  I: Dust cosmologies},
  {\em Gen. Rel. Grav.} {\bf 32} (2000) 105 [{\tt gr-qc/9906015}];
  {\em On average properties of inhomogeneous fluids in general
   relativity:  Perfect fluid cosmologies}, {\em Gen. Rel. Grav.} {\bf
    33} (2001) 1381 [{\tt gr-qc/0102049}];
  {\em Dark Energy from Structure - A Status Report}, {\em
    Gen. Rel. Grav.} {\bf 40} (2008) 467 [{\tt arXiv:0707.2153}]
\bibitem{backreaction} D.J.~Schwarz, {\em Accelerated expansion
  without dark energy} in {\em On the nature of dark energy}, edited
  by P.~Brax, J.~Martin and J.-P.~Uzan (Frontier Group, Paris, 2002)
  p.~331 [{\tt astro-ph/0209584}]; S.~R\"as\"anen, {\em Dark energy
    from backreaction}, {\em JCAP} {\bf 0402} (2004) 003 [{\tt
      astro-ph/0311257}]; E.W.~Kolb, S.~Matarrese, A.~Notari and
  A.~Riotto, {\em Effect of inhomogeneities on the expansion rate of
    the Universe}, {\em Phys.Rev. D} {\bf 71} (2005) 023524 [{\tt
      hep-ph/0409038}]; E.W.~Kolb, S.~Matarrese and A.~Riotto, {\em On
    cosmic acceleration without dark energy}, {\em New J.Phys.} {\bf
    8} (2006) 322 [{\tt astro-ph/0506534}]; A.~Paranjape and T.P.~Singh,
  {\em The Spatial Averaging Limit of Covariant Macroscopic Gravity -
    Scalar Corrections to the Cosmological Equations}, {\em
    Phys.Rev.D} {\bf 76} (2007) 044006 [{\tt gr-qc/0703106}];
  N.~Li and D.J.~Schwarz, {\em On the onset of cosmological
    backreaction}, {\em Phys.Rev. D} {\bf 76} (2007) 083011 [{\tt
      gr-qc/0702043}]; 
  R.A.~Vanderveld, E.E.~Flanagan and I.~Wasserman, {\em Systematic
    corrections to the measured cosmological constant as a result of
    local inhomogeneity}, {\em Phys.Rev. D} {\bf 76} (2007) 083504
  [{\tt arXiv:0706.1931}]; 
  C.-H.~Chuang, J.-A.~Gu and W.-Y.P. Hwang,
  {\em Inhomogeneity-Induced Cosmic Acceleration in a Dust Universe},
  {\em Class.Quant.Grav.} {\bf 25} (2008) 175001 [{\tt
      astro-ph/0512651}]; J.~Behrend, I.A.~Brown and G.~Robbers, {\em
    Cosmological Backreaction from Perturbations}, {\em JCAP} {\bf
    0801} (2008) 013 [{\tt arXiv:0710.4964}]; 
  N.~Li, M.~Seikel and D.J.~Schwarz, {\em Is dark energy an effect of
    averaging?}, {\em Fortsch.Phys.} {\bf 56} (2008) 465 [{\tt
      arXiv:0801.3420}] 
  D.L.~Wiltshire, {\em
    Dark energy without dark energy} in {\em Dark Matter in
    Astroparticle and Particle Physics: Proceedings of the 6th
    International Heidelberg Conference} edited by
  H.V.~Klapdor-Kleingrothaus and G.F.~Lewis (World Scientific,
  Singapore, 2008) p.~565 [{\tt arXiv:0712.3984}]; J.~Larena {\em et
    al.}, {\em Testing backreaction effects with observations}, {\em
    Phys. Rev. D} {\bf 79} (2009) 083011 [{\tt arXiv:0808.1161}];
  C.~Clarkson, K.~Ananda and J.~Larena, {\em The influence of
    structure formation on the cosmic expansion}, {\em Phys.Rev. D}
  {\bf 80} (2009) 083525 [{\tt arXiv:0907.3377}]
\bibitem{li} N.~Li and D.J.~Schwarz, {\em Scale dependence of
    cosmological backreaction},  {\em Phys. Rev. D} {\bf 78} (2008)
    083531 [{\tt arXiv:0710.5073}]
\bibitem{komatsu} E.~Komatsu {\em et al.}, {\em Five-Year Wilkinson
  Microwave Anisotropy Probe (WMAP) Observations:Cosmological
  Interpretation}, {\em Astrophys. J. Suppl.} {\bf 180} (2009) 330
  [{\tt arXiv:0803.0547}]
\bibitem{hicken} M.~Hicken {\em et al.}, {\em Improved Dark Energy
  Constraints from $\sim$100 New CfA Supernova Type Ia Light Curves}, {\em
  Astrophys. J.} {\bf 700} (2009) 1097 [{\tt arXiv:0901.4804}]
\bibitem{guy} J.~Guy {\em et al.}, {\em SALT2: using distant
  supernovae to improve the use of Type Ia supernovae as distance
  indicators}, {\em Astron.Astrophys.} {\bf 466} (2007) 11 [{\tt
    astro-ph/0701828}]  
\end{thebibliography}
\end{document}